\documentclass[authorversion,nonacm,sigconf,natbib]{acmart}
\pdfoutput=1

\usepackage{adjustbox}
\usepackage{csquotes}
\usepackage{enumitem}
\usepackage{listings}
\usepackage{multirow}
\usepackage[caption=false,font=footnotesize]{subfig}
\usepackage{tikz}
\usepackage{pgfplots}
\usepackage{pgfplotstable}
\usepackage{xspace} 
\usepackage{microtype} 
\microtypesetup{expansion=alltext,step=1} %

\pgfplotsset{compat=1.16}
\usetikzlibrary{patterns}

\definecolor{cred}{HTML}{CA0020}
\definecolor{corange}{HTML}{F4A582}
\definecolor{clightblue}{HTML}{92C5DE}
\definecolor{cdarkblue}{HTML}{0571B0}

\pgfplotscreateplotcyclelist{myline}{
    {color=cred,mark=o},
    {color=cdarkblue,mark=star}
}

\pgfplotscreateplotcyclelist{mybar}{
    {fill=cred},
    {fill=clightblue},
    {fill=corange},
    {fill=gray}
}

\usepackage[capitalise,nameinlink]{cleveref}
\crefname{lstlisting}{\lstlistingname}{\lstlistingname}
\Crefname{lstlisting}{Listing}{Listings}

\usepackage{hyperref}

\lstdefinestyle{mystyle}{
    stringstyle=\color{purple},
    basicstyle=\ttfamily\small,
    breakatwhitespace=false,
    breaklines=true,
    captionpos=b,
    keepspaces=true,
    numbers=left,
    numbersep=5pt,
    showspaces=false,
    showstringspaces=false,
    showtabs=false,
    tabsize=2
}

\setcopyright{acmcopyright}
\copyrightyear{2021}
\acmYear{2021}
\acmDOI{10.1145/1122445.1122456}

\acmConference[Woodstock '18]{Woodstock '18: ACM Symposium on Neural
  Gaze Detection}{June 03--05, 2018}{Woodstock, NY}
\acmBooktitle{Woodstock '18: ACM Symposium on Neural Gaze Detection,
  June 03--05, 2018, Woodstock, NY}
\acmPrice{15.00}
\acmISBN{978-1-4503-XXXX-X/18/06}

\usepackage[disable]{todonotes}

\newcommand{\eg}{ e.\,g.,\ }
\newcommand{\ie}{ i.\,e.,\ }

\newcommand{\first}{\emph{(i)}\xspace}
\newcommand{\second}{\emph{(ii)}\xspace}
\newcommand{\third}{\emph{(iii)}\xspace}

\newcommand{\ohalldecnormal}{104.66}
\newcommand{\ohallencnormal}{80.50}
\newcommand{\ohhttpsdecnormal}{93.56}
\newcommand{\ohhttpsencnormal}{65.51}
\newcommand{\ohwebdecnormal}{89.42}
\newcommand{\ohwebencnormal}{71.29}

\newcommand{\ohalldecmod}{104.53}
\newcommand{\ohallencmod}{237.91}
\newcommand{\ohhttpsdecmod}{91.62}
\newcommand{\ohhttpsencmod}{215.41}
\newcommand{\ohwebdecmod}{90.59}
\newcommand{\ohwebencmod}{213.51}

\newcommand{\patchurl}{\url{https://github.com/UHH-ISS/zeek}}
\newcommand{\pcapurl}{\url{https://mega.nz/folder/MgQRESgA\#5jYh37bZ3MKZqp7yeVemgQ}}

\begin{document}

\title{Passive, Transparent, and Selective TLS Decryption\\ for Network Security Monitoring}

%
\author{Florian Wilkens}
\affiliation{%
  \institution{Universität Hamburg}
  \state{Hamburg}
  \country{Germany}
}
\email{wilkens@informatik.uni-hamburg.de}

\author{Steffen Haas}
\affiliation{%
  \institution{Universität Hamburg}
  \state{Hamburg}
  \country{Germany}
}
\email{haas@informatik.uni-hamburg.de}

\author{Johanna Amann}
\affiliation{%
  \institution{ICSI}
  \state{California}
  \country{United States}
}
\email{johanna@icir.org}

\author{Mathias Fischer}
\affiliation{%
  \institution{Universität Hamburg}
  \state{Hamburg}
  \country{Germany}
}
\email{mfischer@informatik.uni-hamburg.de}

\renewcommand{\shortauthors}{Wilkens et al.}

\begin{abstract}
Internet traffic is increasingly encrypted. While this protects the confidentiality and integrity
of communication, it prevents network monitoring systems~(NMS) and intrusion detection systems
(IDSs) from effectively analyzing the now encrypted payloads.
Therefore, many enterprise networks have deployed man-in-the-middle~(MitM) proxies that intercept
TLS connections at the network border to examine packet payloads and thus retain some visibility.
However, recent studies have shown that TLS interception often reduces connection security and
potentially introduces additional attack vectors to the network.
%
In this paper, we present a cooperative approach in which end-hosts as cryptographic endpoints
selectively provide TLS key material to NMS for decryption. This enables endpoints to control who
can decrypt which content and lets users retain privacy for chosen connections.
We implement a prototype based on the Zeek NMS that is able to receive key material from hosts,
decrypt TLS connections and perform analyzes on the cleartext. The patch is freely available and we
plan to upstream our changes to Zeek once they are mature enough.
%
In our evaluation, we discuss how our approach conceptually requires significantly less computational
resources compared to the commonly deployed MitM proxies. Our experimental results indicate, that TLS
decryption increases a runtime overhead of about 2.5 times of the original runtime on cleartext.
Additionally, we show that the latency for transmitting keys between hosts and the NMS can
be effectively addressed by buffering traffic at the NMS for at least 40ms, allowing successful
decryption of 99.99\% of all observed TLS connections.
\end{abstract}

\settopmatter{printfolios=true}
\maketitle

\section{Introduction}\label{sec:introduction}
Nowadays, the majority of Internet traffic is encrypted, and the biggest share of it by Transport
Layer Security~(TLS). For example, Google reports that about 90\% of all websites visited by their
Chrome Browser use HTTPS\footnote{\url{https://transparencyreport.google.com/https/overview}}.
However, TLS prevents the analysis of packet payloads by network monitoring systems~(NMS) and
restrict them to connection metadata. While some systems can work with NetFlow data
only~\cite{anderson2016identifying}, most require an inspection of the payload,\ie{}deep packet
inspection~\cite{antonello2012deep}. Especially in the context of security monitoring, access
to cleartext payloads is essential either for real-time intrusion detection or for forensic
analysis as part of incident response. While flow directions are helpful to reconstruct the
attacker's movement in the network, payload data can help to identify attack vector and
tools the attacker used throughout the incident.

In enterprise network administrators often deploy mandatory man-in-the-middle~(MitM) proxy servers
that terminate TLS at the organization boundary to enable effective security monitoring. These
proxies then forward decrypted traffic to NMSs for cleartext payload analysis.
However, terminating TLS connections raises several concerns:
\first The authentication mechanism via server certificates is effectively circumvented as the
client is required to fully trust the proxies' certificate authority (CA). An attacker, that manages
to establish a MitM position in the network and compromise this CA, could intercept and modify any
encrypted traffic unnoticed.
\second The MitM proxy raises privacy concerns for users as all traffic is intercepted without
exception. While this might be reasonable in high-security environments, a regular enterprise
network should let users retain privacy for domains that are unlikely to endanger the network.
\third The MitM proxy adds another potential attack vector to the network that needs to be patched
and maintained. This is especially important as the proxy by definition has complete access to the
traffic and could arbitrarily modify it. Additionally, recent studies have confirmed, that
intercepting TLS sessions often weakens the overall connection
security~\cite{Jarmoc2012,de2016killed,Durumeric2017,o2016tls,o2017tls,waked2018intercept}.

To address these shortcomings, a solution is required that allows the NMS to process TLS
application data in cleartext without intercepting the TLS connection and by upholding the
end-to-end integrity protection between both communication partners.

The main contribution of this paper is a generic system for the passive inspection of TLS encrypted
traffic without the need for a MitM proxy. Instead, we suggest that hosts forward TLS key material
to a NMS, such that it can directly decrypt and analyze packets in TLS sessions without taking an
active role in the connections.
In contrast to other state-of-the-art solutions like proxies or TLS extensions for multi-entity
support~\cite{badra2011securing}, our approach:
\begin{itemize}[noitemsep,topsep=6pt]
    \item works without modifications to the TLS protocol,
    \item adds no latency to TLS session establishment,
    \item enables only trusted observers to inspect TLS sessions,
    \item allows for selective traffic inspection controlled by endpoints themselves,
    \item and can preserve the end-to-end integrity of the TLS session for non-AEAD cipher suites
        as the observer only receives decryption but no integrity keys.
\end{itemize}

We implement our approach as a prototype for the open-source Zeek NMS (formerly known as
Bro)~\cite{paxson1999bro}. Our prototype extends Zeek to be able to receive key material from
endpoints and use it to decrypt TLS sessions. The resulting application data payloads are
transparently handed over to Zeek's builtin analyzers for inspection. We intend to upstream our
patch to Zeek once it is mature enough and it is available on github for now\footnote{\patchurl}.

We evaluate our prototype along two different dimensions: \first We measure the \emph{decryption
overhead} by comparing the computational overhead for processing encrypted and cleartext traffic.
\second We estimate the \emph{impact of transfer latency of key material} between client and NMS on
the amount of decrypted packets, bytes and number of connections.
Our result indicate that decryption imposes a 2.5 times computational overhead compared to the
processing of the same, but not encrypted traffic. In real-world deployments, the latency for
transferring keys is rather low. It requires only a small traffic buffer in front of the NMS
Results from our evaluation indicate that 40ms of buffering time are sufficient to decrypt 99.99\%
of observed TLS connections in our testbed.

The remainder of the paper is structured as follows. \cref{sec:bg_sota} summarizes some background
on TLS, states objectives for our approach and reviews the state of the art. \cref{sec:system}
presents the design of our approach that enables TLS decryption without having to terminate the
connection. \cref{sec:evaluation} briefly discusses the computational complexity of our approach
compared to MitM proxies, describes our prototype and testbed and evaluates our implementation on a
captured data set using the Alexa Top 1000 before our paper concludes in \cref{sec:conclusion}.

\section{Objectives \& State of the Art}\label{sec:bg_sota}
In this section, we briefly recap some TLS fundamentals and describe objectives for passive and
transparent TLS decryption on a NMS. Additionally, we summarize and discuss the state of the art in
the areas of MitM proxies, TLS protocol modifications, and trusted execution environments along our
objectives.

\begin{figure*}
    \subfloat[Full handshake\label{fig:tls12-handshake-full}]{%
        \includegraphics[width=.48\linewidth]{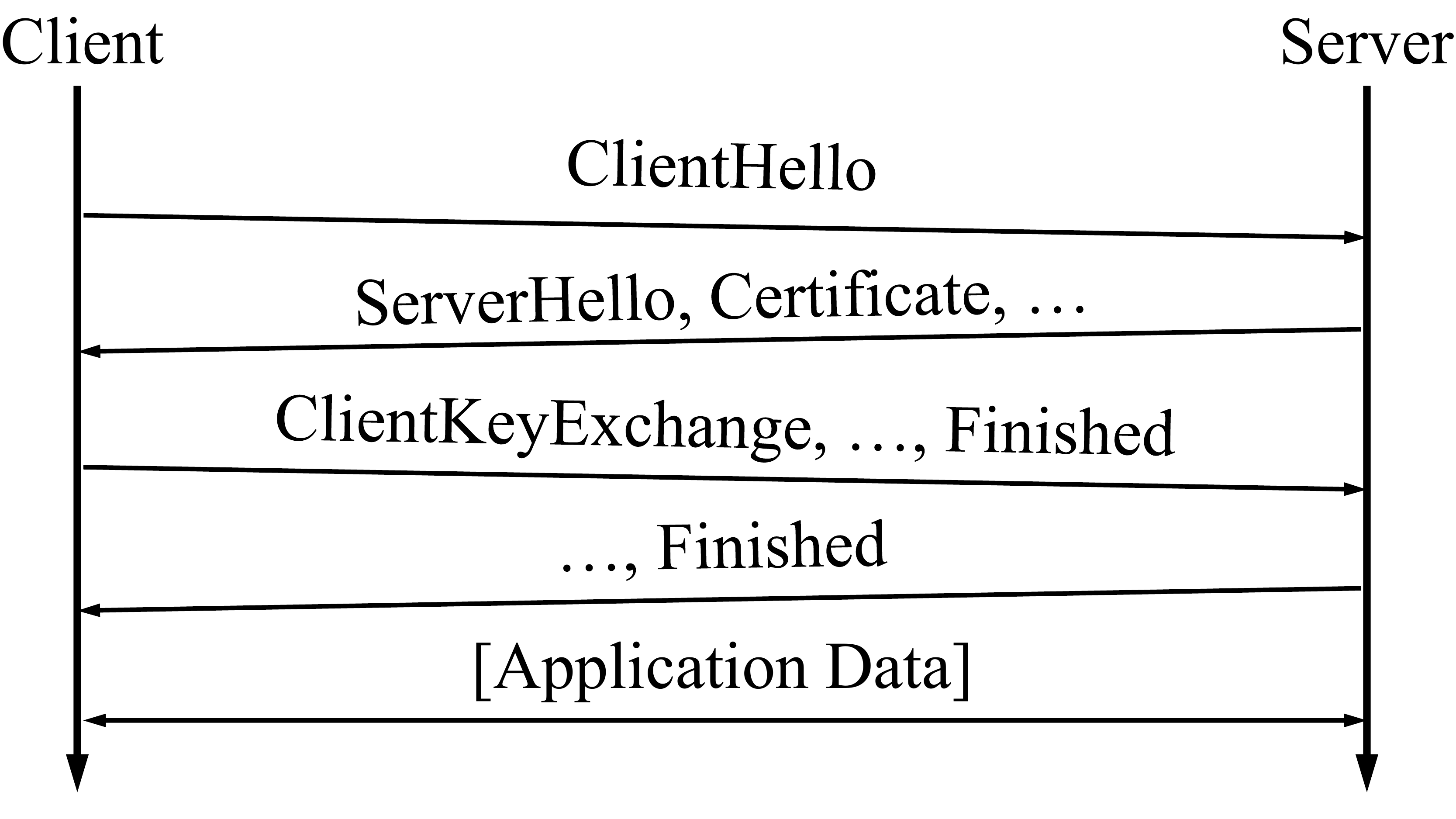}
    }
    \hfill
    \subfloat[Resumed handshake via SessionID\label{fig:tls12-handshake-resumed}]{%
        \includegraphics[width=.48\linewidth]{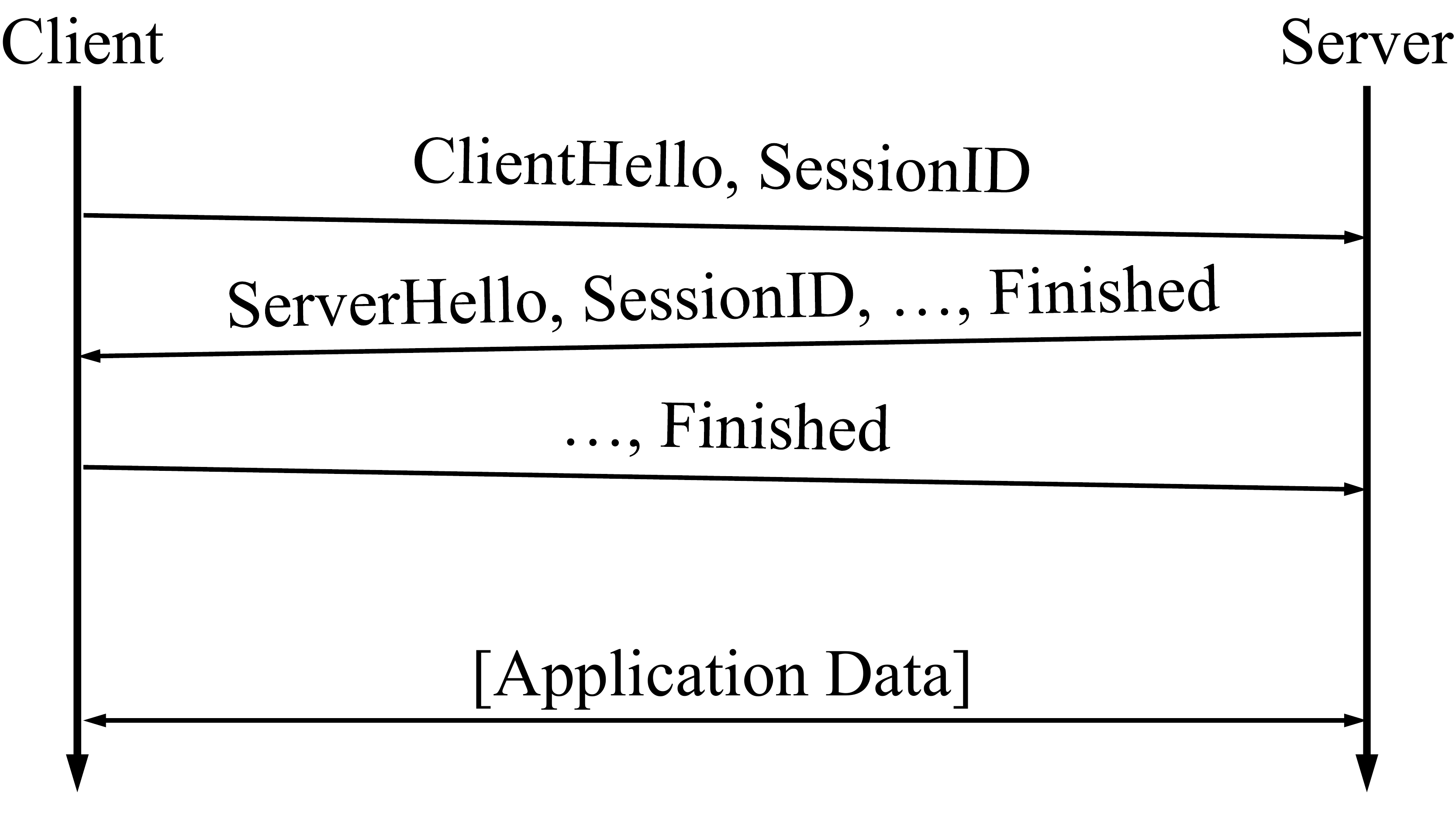}
    }
    \caption{TLS 1.2 handshakes}\label{fig:tls12-handshakes}
\end{figure*}

\subsection{TLS Fundamentals}\label{subsec:bg_sota-tls}
Transport Layer Security~(TLS) was designed to protect confidentiality and data integrity between
two communication partners. This section briefly summarizes the TLS protocol details relevant to
our contribution.
Note: As of 2021 all version below TLS 1.2 have been deprecated~\cite{Rfc8996}, so that we will
only focus on TLS 1.2 and 1.3.

All TLS connections start with a handshake between the two endpoints. In this handshake the
endpoints agree on a cipher suite to protect application data, and establish the necessary
cryptographic material.
To authenticate, the endpoints can also verify each other's identity via X.509 certificates. During
a standard TLS handshake only the server authenticates towards the client, while the client
authenticity is checked by other means,\eg{}via entering user credentials on the website retrieved.

\Cref{fig:tls12-handshakes} shows two variants of a TLS 1.2 handshake:
\Cref{fig:tls12-handshake-full} depicts the full handshake that is performed for a new connection
without any preceding communication. As no prior information is available both endpoints negotiate
a cipher suite for the connection and perform a key exchange to establish shared cryptographic
material, called the \emph{pre-master secret}.
\Cref{fig:tls12-handshake-resumed} shows one variant of a resumed handshake via \emph{session ID}s.
In this case, both communication endpoints reuse cryptographic material from a prior connection and
thus skip the key exchange part of the handshake. TLS 1.2 also offers session resumption on the
basis of a \emph{session ticket} that contains all cryptographic state to resume a session. To
relieve the server from storing this material as with session IDs the server encrypts and transmits
it to the client, that then presents it in the resumed handshake.

TLS 1.3 changes both the full and resumed handshake. The full handshake is reduced to one roundtrip
by combining and rearranging selected parts of the handshake such that less roundtrips are
required. Resumption is supported through a modified session ticket approach based on pre-shared
keys~(PSK). However, the concept remains unchanged: \first in full handshakes a key exchange scheme
is used to establish a shared secret. \second resumed handshakes skip this and used established key
material in form of a PSK. However, the PSK is altered for every new connection and sent encrypted
compared to session IDs and tickets that are both sent in the clear.

After a shared secret has been established, integrity and confidentiality keys are derived by both
endpoints. \Cref{lst:tls12-key-derivation} shows this \emph{key derivation} process for TLS 1.2.
First, a \emph{master secret} is derived by using the TLS pseudo-random function~(PRF) as defined
in RFC 5246~\cite{Rfc5246}. This function uses a hash function depending on the selected cipher
suite. Next, the PRF is used again to derive up to three sets of secrets containing one secret for
server and client, respectively. The encryption keys (\texttt{client\_key}, \texttt{server\_key})
are always used. The integrity of the encrypted data is either ensured via a separate hash-based
message authentication code~(HMAC) or by using an authenticated encryption with associated
data~(AEAD) cipher that integrates data integrity checks into the decryption process.
In the first case, \texttt{client\_MAC} and \texttt{server\_MAC} are used to key the HMAC. For AEAD
ciphers, the HMAC keys are derived and discarded (to retain the order of derived secrets) and two
initialization vectors~(IVs) (\texttt{client\_IV}, \texttt{server\_IV}) are derived after the encryption
keys. This is simple as the PRF can be used to derive any number of bytes.

\begin{lstlisting}[
    style=mystyle,
    caption={TLS 1.2: Key derivation from pre-master secret},
    label={lst:tls12-key-derivation},
    language=c
]
pre_master_secret = <obtained from key exchange>
master_secret = TLS_PRF(
    pre_master_secret,
    "master secret",
    client_random + server_random)
key_buffer = TLS_PRF(
    master_secret,
    "key expansion",
    server_random + client_random);

client_MAC = key_buffer[0..31]
server_MAC = key_buffer[32..63]
client_key = key_buffer[64..79]
server_key = key_buffer[80..95]
client_IV = key_buffer[96..99]
server_IV = key_buffer[100..103]
\end{lstlisting}

Once keys for encryption and integrity protection are available, all following data in the TLS
connection is both integrity and confidentiality protected using the negotiated cipher suite.

\subsection{Objectives for passive TLS Decryption}\label{subsec:bg_sota-requirements}
Before we summarize state of the art in the next section, we briefly describe objectives for the
goal of passive TLS decryption already outlined in~\cref{sec:introduction}. These objectives are
used to evaluate both related work and our own concept and prototype described later.
To regain visibility into encrypted TLS traffic on a trusted NMS, a well-designed solution should
exhibit the following properties:
%
%
\begin{itemize}
    \item \textbf{No interception}: TLS is a well-designed protocol to ensure end-to-end data
        confidentiality and integrity between two endpoints. Additional entities, that are inserted
        into the connection to intercept it, represent new and unwanted attack vectors that may
        weaken the security of the connection.
        Thus, a proper solution should not intercept the TLS connection.
    \item \textbf{No protocol modification}: TLS is deployed widespread across a wide variety of
        devices and networks. Thus, changes on the protocol-level are unlikely to gain enough
        traction for widespread adoption. To strive for real-world usage, a solution should not
        require any changes to the existing TLS standards.
    \item \textbf{No additional latency}: While security and network monitoring is an important
        topic in any enterprise network, it should not detriment the user experience by introducing
        additional latency to the connection.
    \item \textbf{Support selective decryption}: Endpoints should be able selectively control which
        connections are allowed to be decrypted by the NMS to let the user retain some privacy for
        sensitive or personal data. 
        Users could be given an opt-out mechanism to prevent that selected TLS connections are not
        touched at all, if that is feasible by network policy.
    \item \textbf{Low maintenance effort}: Security monitoring is an essential service in an
        enterprise network and as such should require low maintenance for regular operations. This
        is both relevant for client machines and any middleboxes deployed in the network.
    \item \textbf{Maintain end-to-end data integrity}: The NMS needs to break end-to-end
        confidentiality between client and server to perform analysis tasks on the cleartext.
        However, this process does not require integrity violation as application data only needs
        to be read by the NMS.
        Thus, a good solution should maintain end-to-end data integrity for inspected TLS
        connections if possible.
\end{itemize}

\subsection{State of the Art}\label{subsec:bg_sota-sota}
\subsubsection{TLS interception via MitM}\label{subsubsec:bg_sota-sota-mitm}
The most popular way to decrypt TLS traffic is to use a man-in-the-middle~(MitM) proxy that
terminates all connections and thus has access to the cleartext data. As industry best-practice
this approach is often deployed in enterprise networks. Many implementations are available
depending on the use-case and inner protocol that should be analyzed including open-source tools
such as \emph{mitmproxy}~\cite{mitmproxy} and sslplit\footnote{\url{https://www.roe.ch/SSLsplit}}.
Commercial offerings for network security also often include a component that intercepts TLS either
as a separate application or directly integrated into the IDS or NMS.

Due to its popularity, TLS interception has been discussed broadly in recent years with the
overwhelming concensus that it significantly weakens the security properties of intercepted
connections~\cite{Jarmoc2012,de2016killed,Durumeric2017,o2016tls,o2017tls,waked2018intercept}.
%
In~\cite{Jarmoc2012} Jarmoc et al. discuss the security of intercepting end-to-end encryption in
general. They state, that there is no implementation solely for the purpose of only intercepting
TLS connections, but rather multiple classes of applications such as web proxies, data loss
prevention system or network IDSs sometimes include an integrated feature for this. According to
the authors all of them introduce potential risk into the environment they protect including legal
peril, an increase in the overall threat surface by having a single point of decryption and a
possible decreased cipher strength among others.
%
More recently, Durumeric et al.~\cite{Durumeric2017} also confirm that intercepting proxies and the
corresponding security degradation for clients are still prevalent. They find that there is an
order of magnitude more interception than previously estimated. Additionally, they evaluate
multiple middleboxes and conclude that nearly all reduce overall connection security and many
introduce severe vulnerabilities.


\subsubsection{Modifications of the TLS protocol}\label{subsubsec:bg_sota-sota-tls}
TLS is designed to support end-to-end encryption between exactly two entities. However, there are
proposals to add multi-entity support that would enable previously defined forwarding endpoints to
decrypt the payload as well.

\begin{table*}
    \centering
    \begin{tabular}{lcccccc}
        \toprule
        Requirement              & MitM proxies & mcTLS/mbTLS~\cite{Naylor2015,Naylor2017} & maTLS~\cite{Lee2019} & ETLS~\cite{Etsi2018} & EndBox~\cite{Goltzsche2018} & ShieldBox~\cite{Trach2018} \\
        \midrule
        No interception          & ---          & ---          & ---          & \checkmark   & \checkmark   & \checkmark   \\
        No protocol modification & \checkmark   & ---          & ---          & (\checkmark) & \checkmark   & \checkmark   \\
        No additional latency    & ---          & ---          & ---          & \checkmark   & \checkmark   & \checkmark   \\
        Selective decryption     & ---          & (\checkmark) & (\checkmark) & (\checkmark) & (\checkmark) & (\checkmark) \\
        Low maintenance effort   & (\checkmark) & ---          & ---          & \checkmark   & ---          & ---          \\
        E2E data integrity       & ---          & (\checkmark) & (\checkmark) & ---          & \checkmark   & \checkmark   \\
        \bottomrule
    \end{tabular}
    \caption{Objectives achieved by state of the art.\label{tb:sota}\\
    ---\ not achieved\hspace*{.5cm}(\checkmark)\ partially achieved\hspace*{.5cm}\checkmark\ achieved}
\end{table*}

In \cite{Naylor2015} the authors propose Multi-Context TLS~(mcTLS) that extends the TLS header by
multiple so-called contexts. These contexts act as a permission system and enable the client to
allow selected middleboxes to read and write the TLS payload. This enables these middleboxes to
perform various in-network services that are not possible with standard TLS including intrusion
detection or more general network monitoring. However, mcTLS has two major downsides: \first the
additional contexts require additional computation on the client and add additional latency during
connection establishment. \second the modified TLS header with contexts is incompatible to standard
TLS and thus forces all nodes on the path to support mcTLS.
%
The second downside is partially addressed in Middlebox TLS~(mbTLS) also by Naylor et
al.~\cite{Naylor2017}. mbTLS extends the standard TLS handshake with a custom extension that
middleboxes on the path can use to inject themselves into the connection with the client's consent.
The endpoints then establish secondary TLS sessions with each of their respective middleboxes. As a
non-standard extension is used, endpoints can fall back to standard TLS if one side does not
support mbTLS. To establish trust in the middleboxes, remote attestation via an enclave is used.
Overall, this improves on mcTLS but further complicates deployment and maintenance as middleboxes
now require a hardware enclave. Furthermore, additional latency is added by the secondary TLS
sessions.

Middlebox-aware TLS~(maTLS)~\cite{Lee2019} modifies TLS, so that all middleboxes on the
communication path explicitly forward security parameters to the client. This is implemented via a
combination of Split TLS with added support of entity checks in a new TLS protocol. It ensures
explicit authentication, security parameter validation to ensure confidentiality and modification
validity checks for integrity. Every middlebox posesses it's own certificate which can be verified
by a third-party. Like mcTLC, this also is incompatible to standard TLS with all the downsides.
Like mcTLS and mbTLS especially the connection establishment requires additional computation time
and delay is caused by the multiple entities that potentially perform computations on the
connection.

Legislators have also pushed to modify TLS encryption in a way that would allow access for law
enforcement. Enterprise TLS~(ETLS)~\cite{Etsi2018} as proposed by the European Telecommunications
Standards Institute~(ETSI) changes the TLS 1.3 standard to make forward secrecy optional and keep
the outdated RSA mode available. In practice service providers could then always use the same key
and thus leave access to authorities, even at some point in the future.

\subsubsection{Trusted Execution Environments}\label{subsubsec:bg_sota-sota-tee}
Trusted execution environment have been used to deploy middlebox functionality like network
monitoring to communication endpoints. As the cleartext is available on these machines, these
approaches do not need to modify the TLS protocol to perform analysis tasks on the payloads.

In~\cite{Goltzsche2018} the authors present EndBox, a solution with a focus on the middlebox
functions instead of connection data. They distribute the middlebox application,\eg{}the NMS, in a
decentralized fashion. The software is completely isolated from the operating system by running in
a trusted execution environment. From a privacy perspective the main advantage is that connection
payloads remain untouched between client and server. However, the implementation effort is high:
The client part of the software needs to be modified in order to run in the enclave. More
importantly, every application on the middlebox that needs to have access to the payload must be
modified in order to be compatible. This is a very restrictive requirement and limits the set
applications. Furthermore, the approach is based on Intel SGX and thus puts extra restrictions on
which hardware can be used on the clients.

Trach et al.~\cite{Trach2018} present ShieldBox that also use Intel SGX. They try to integrate the
middlebox functionality into the trusted environment on the server. The goal is the ability to
outsource middlebox applications to remote untrusted platforms. To be able to use using existing
software, they designed a standard C library (libc) called SCONE that supports running in an Intel
SGX environment. The authors enhanced the framework to support fast processing of network packets.
However, this approach also has some downsides: there are major limitations on memory usage and
many operations (especially system calls) are consume more time. A general usage of open-source
software cannot be assumed. Additionally, the same high installation and maintenance challenges as
for Endbox remain.

\subsubsection{Summary}
\Cref{tb:sota} gives an overview about the state of the art and indicates which of our objectives
are fulfilled by the respective approach. MitM proxies offer the advantage, that the TLS protocol
does not need to be modified and maintenance effort is relatively low as only the proxy itself
needs to be maintained and the CA needs to be deployed to clients. However, dues to it's MitM
position, the proxy gains complete access to all cleartext payloads breaking the E2E data integrity
and adding additional latency. While selective decryption could be implemented it can not be
conceptually guaranteed. mcTLS and mbTLS modify the TLS protocol to support middleboxes. Their
permission system could be used to implement selective decryption and maintain the E2E data
integrity, however it is not supported out of the box. The maintenance effort is rather high, as
middleboxes and respective endpoints need to be kept in sync due to the protocol. maTLS has some
slightly different goals compared to mcTLS/mbTLS but checks the same of our objectives. Selective
decryption and E2E integrity could potentially be implemented but are not the focus. ETLS checks
most of our objectives but has a few key flaws. While the protocol is not substantially modified,
key security features such as perfect forward secrecy are intentionally disabled to allow external
audits through a shared private key. While this could also be seen as some kind of selective
decryption, it does not retain the user in control of which connections are decryptable and rather
adds another attack vector as the private key that is potentially shared with third parties could
be compromised. EndBox and ShieldBox both employ Intel SGX to move NMS functionality to the
endpoints. This has many upsides, as the protocol is not modified and no additional middleboxes
are inserted into the connection. However, two major downsides remain: \first deployment and
maintenance overhead increase as the endpoints need to be equipped with enclave hardware. \second
the NMS on the endpoint conceptually has access to all cleartext data of the connections. While
selective decryption can be implemented in the enclave, users have to rely on the correctness of
the implementation and cannot prevent the NMS from analyzing selected connections.

In summary, no existing approach fulfills all our objectives. We also note, that true selective
decryption can only be achieved, if the client somehow retains control over the TLS key material.
Thus, our approach is based on cooperative endpoint that selectively forward key material to a
passive NMS that then decrypts and analyzes the cleartext payloads. We describe this approach in
the next section.

\section{Passive TLS Decryption at a NMS}\label{sec:system}
\begin{figure*}
    \centering
    \includegraphics[width=.7\linewidth]{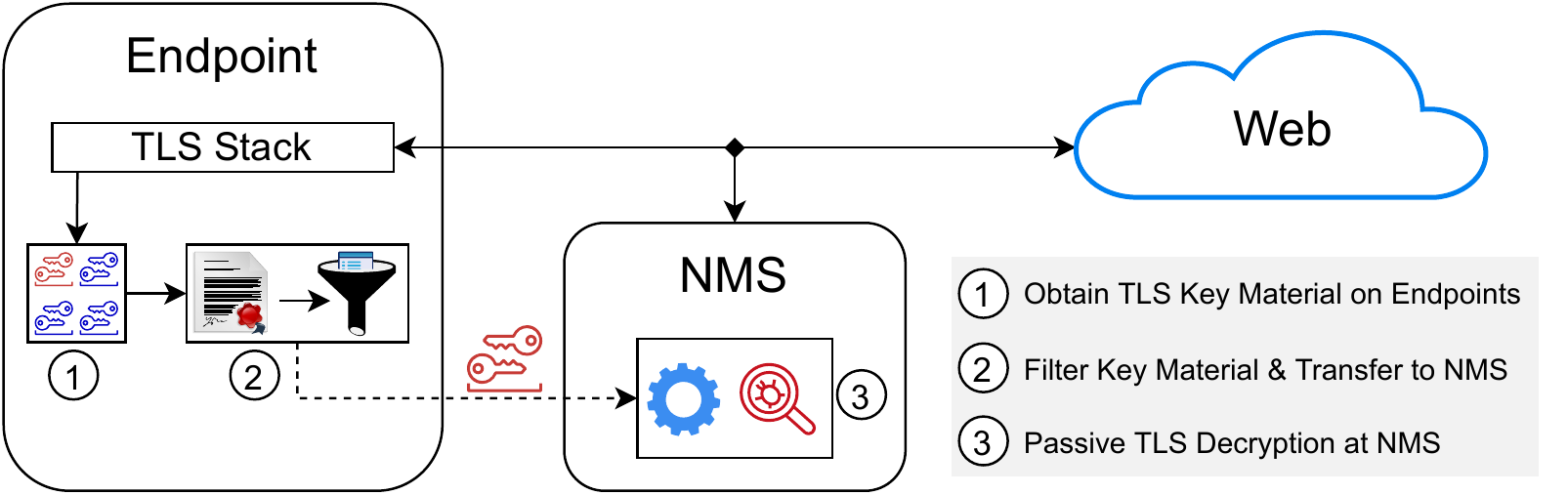}
    \caption{Overview on passive TLS Decryption via cooperative endpoints}\label{fig:overview}
\end{figure*}

In this chapter, we describe the concept of passively decrypting TLS connections at a network
monitoring system~(NMS).
\Cref{fig:overview} gives an overview about the complete process and components involved. First,
the endpoints in the network protected by the NMS need to obtain the TLS session keys. Once they
have been obtained, they need to be transferred to the NMS. Finally, the NMS matches the key
material to connections and starts passive decryption and further analysis.
%

While this approach seems to be conceptually simple, the individual steps encompass conceptual and
technical challenges that we describe in the following sections.

\subsection{Obtain TLS Session Keys on Endpoints}\label{subsec:system-obtain}
In the first step, TLS key material needs to be obtained on the client machines as they represent
the cryptographic endpoint that is usually under control in an enterprise network. However, a
typical TLS connection, uses multiple secrets as well as session keys for different purposes as
described in~\cref{subsec:bg_sota-tls}. The simplest solution would be to obtain the pre-master
secret and forward it to the NMS with the client random as the connection identifier. The NMS would
then perform key derivation as shown in~\cref{lst:tls12-key-derivation} and obtain all session keys
(and IVs if an AEAD cipher suite is used) to decrypt the encrypted application data packets.


However, this enables the NMS to derive all session keys including the integrity keys. This would
allow a compromised NMS to forge packets on behalf of both endpoints and thus introduce an
additional attack vector into the network.
For non-AEAD ciphers, we can mitigate this by only forwarding encryption keys to the NMS and
keeping the integrity keys used by the HMAC on the client. Unfortunately, AEAD ciphers require both
encryption keys and IVs to decrypt the data even if no integrity checks need to be performed by the
NMS.

Once the decision has been made, which key material will be forwarded it needs to be collected on
the client.
The key material can be obtained in multiple ways depending on operating system, cryptographic
library, and application. The simplest way is the \textbf{SSL keylog interface} supported by widely
used cryptographic libraries such as openSSL and NSS~\cite{mdn-keylog}. If enabled, the libraries
write TLS pre-master secrets and the corresponding client randoms to the file specified in an
environment variable. While this approach is easy to deploy, it might introduce some problems:
\first The logfile is readable by the user, as the browser process is running with user
permissions. \second Writing secrets to a file and reading them in the aftermath via an application
adds additional latency. While this can be partially addressed by using a RAM disk, the NMS might
receive the keys too late and the first few packets of a TLS connection cannot be decrypted.

The \textbf{kernel TLS (kTLS)} interface~\cite{Watson2016} offers another way to obtain the desired
key material. kTLS was introduced to the Linux kernel in version 4.13 and moves the TLS packet
handling to kernel space. At the time of writing only a limited subset of AEAD cipher suites is
supported. To use it, an application opens a regular socket and performs the usual TLS setup via
key exchange and derivation. Next, it configures the encryption key as well as IV for the socket
via a \texttt{setsockopt} system call. From that point on, regular \texttt{send} and
\texttt{receive} operations can be used with the kernel performing encryption and integrity checks.
A process with elevated permissions can monitor this system call for keys, extract them and forward
them to the NMS.

\subsection{Filter Key Material \& Transfer to NMS}\label{subsec:system-transfer}
Once the key material has been obtained on the client, it has to be transferred to the NMS together
with a unique connection identifier.
The actual transmission method used is dependent on the NMS in question. As key material is highly
sensitive, its transmission should be properly secured,\eg{} by using mutually authenticated TLS
connections between endpoints and NMS.

When endpoints are involved this also raises the question if the NMS has to gain visibility into all
TLS connections or if exceptions can be made.
In regular enterprise networks, the decryption of all TLS encrypted traffic might not be required.
Users should be able to retain privacy for domains that are unlikely to be used by an attacker,
even though you can never anticipate the actual attack vector used in the end.
As our approach is based on the cooperation of endpoints, these can simply avoid to share key
material for certain TLS connection with the NMS. In such a \emph{selective decryption} the user
could be involved via a white- or blacklist, which controls which keys are actually forwarded to
the NMS. The NMS can then in the end only decrypt packets on selected TLS connections and would fall
back to metadata-only analysis for all others.

However, it is crucial to ensure that the attacker cannot exploit this mechanism by manipulating
the respective black- or whitelists. This requires to protect and check the integrity of such
lists, e.g., by signing them on another machine. The concrete realization of this user involvement
is beyond the scope of this paper and left for future work.
In high-security environments such as governmental institutions or critical infrastructures, our
approach would simply forward all key material to the NMS In this case, all connections are
decrypted, which resembles the classic MitM setting.

\subsection{Passive TLS Decryption at the NMS}\label{subsec:system-decryption}
Once the NMS receives TLS key material from the endpoint, it needs to identify the connection the key
material belongs to. This identifier differs for TLS versions and different implementations of session
resumption. To identify a connection that was established via a full TLS 1.2 handshake, as shown in
\cref{fig:tls12-handshake-full}, the \emph{client random} that is sent with the \emph{ClientHello}
uniquely identifies the connection and thus can be used to match key material. For subsequent resumed
connections, the NMS then needs to also associate the key material with the session identifier. This
is either \emph{session ID}, \emph{session ticket} or \emph{TLS 1.3 PSK} depending on TLS version and
implementation. If the session identifier is then encountered in a resumed handshake, the NMS can match
the key material as for full handshakes. In TLS 1.2 this process is trivial as both session ID and
session ticket are sent in cleartext and thus can be easily stored. The TLS 1.3 PSK approach is slightly
more complicated as the PSK is already encrypted when sent from the server in the first connection and
additionally exchanged after the first use in the resumed connection. However, this does not pose a
problem to our concept as the NMS can already decrypt the first connection with key material from the
client and thus also store the encrypted PSK for later use.

Once the connection has been identified, the NMS uses the obtained secrets to decrypt TLS application
data packets in the connection. This process poses two major challenges:
\first \textbf{Cipher diversity:} The different TLS versions support varying cipher suites that the NMS
needs to implement. This can be addressed by reusing popular cryptographic libraries such as OpenSSL or
NSS. An incremental approach with a subset of supported cipher suites is also possible starting with
popular ciphers.
\second \textbf{Missing key material}: Key material might not be available, once the first encrypted
packet of a TLS connection arrives. This can happen either because the client chose to not send the
key material due to filtering or because the key material is simply still in transit to the NMS. In the
first case, the NMS should simply skip decryption, while the second case would require the NMS to buffer
packets until the key material arrives. However, this might introduce a denial of service attack vector
if the NMS needs to buffer too many packets for connections especially as the NMS cannot know in advance
if key material will arrive at some point in the future. To solve this problem, a traffic buffer in
front of the NMS should be deployed that delays only the encrypted traffic to analyze for a short time
before it is forwarded to the NMS. This way keys can arrive in time for complete decryption. We quantify
the impact of the key transfer latency on the decryption and recommend a buffer size in our evaluation
(c.f.~\cref{subsec:eval-experiment2}).

\section{Evaluation}\label{sec:evaluation}
In this section, we first discuss the computational complexity our approach induces and
compare it to the commonly deployed TLS intercepting MitM proxy. Next, we describe the features and
limitations of the proof-of-concept prototype we implemented to test and evaluate our concept. Finally, we describe
the results from two experiments in which we measured the \emph{decryption overhead} and the \emph{impact of key transfer latency} on the decryption success rate, respectively.

\subsection{Computational Complexity}\label{subsec:eval-computation}
We compare the computational overhead of our approach on a conceptual level with intercepting MitM
proxies. TLS adds non-negligible computation overhead to every connection and the proxy has to
setup and maintain two TLS connections per endpoint connection. A decrypting NMS remains passive
and thus does not directly inititate TLS connections. Instead, it only decrypts the encrypted
payloads to gain access to cleartext data.

\begin{table}
    \centering
    \begin{tabular}{lcccc}
        \toprule
        Approach                  & CV & KE & \(E_{sym}\) & \(D_{sym}\) \\
        \midrule
        TLS Client                & 1  & 1  & \(s\)        & \(r\) \\
        TLS Server                & 0  & 1  & \(r\)        & \(s\) \\
        Intercepting Proxy        & 1  & 2  & \(s+r\)      & \(s+r\) \\
        Decrypting \& passive NMS & 0  & 0  & --           & \(s+r\) \\
        \bottomrule
    \end{tabular}
    \caption{Simplified computational complexity for typical TLS connections}\label{tb:complexity}
\end{table}

\Cref{tb:complexity} summarizes the simplified computational costs of establishing TLS connections
for the TLS client and server respectively, but also for a MitM proxy and a passive NMS. We assume
the default behavior in HTTPS where the client authenticates the server only. The TLS client
performs a \emph{certificate validation} (CV) for the server certificate and a \emph{key exchange}
(KE) to establish session keys. The exact number of cryptographic operations here differs based on
the cipher suite used, but usually involves CPU intensive asymmetric cryptography. Once the key
material is established, the client encrypts the \(s\) bytes of its request via the symmetric
cipher \(E_{sym}\) that was selected in the handshake and sends them to the server. After a
response is received, the \(r\) number of bytes are decrypted. The TLS server performs a similar
number of operations, but usually without certificate validation as TLS clients rarely use
certificates in HTTPS. The number of encrypted and decrypted bytes are switched as request bytes
from clients \(s\) have to be decrypted and response bytes \(r\) encrypted. However, as a symmetric
cipher is used with identical costs of encryption and decryption, the overall computation effort
should be nearly identical.

The MitM proxy acts as TLS server to the client and as the client for the requested TLS server and
thus performs the combined computations of both endpoints. While the KE results in a predictable
amount of computation, \(s\) and \(r\) vary in different scenarios. For small connections such as
simple websites, the doubled amount of symmetric cryptography is less impactful, while large
connections such as a large file download via HTTPS introduce significant amounts of additional
computation on the proxy.
%
The decrypting NMS in our approach has one large conceptual advantage to the proxy: due to its
passive analysis, it only needs to decrypt all bytes \(s+r\) of a single connection once to achieve
cleartext access to the payloads. As no TLS connection is actively established, no KE needs to be
performed and even validation of the server certificate is not strictly required as the client
would rightfully reject the connection in case of failure. However, analysis of the server
certificate might be interesting as a regular NMS analysis outside of decryption.

In summary, our approach conceptually requires less computational intensive operations when compare
with the commonly deployed approach of TLS intercepting proxy servers. While symmetric cryptography
is continously becoming faster to execute on modern hardware due to special instruction sets, it is
still represents the largest share of computation in a proxy and thus dictates how machine
resources need to be scaled for high-bandwidth deployments. Our approach requires significantly
less computations while still enabling the NMS to perform analysis on cleartext payloads. However,
the lowered computational load is shifted from proxy machines to NMS machines. Thus, our first
experiment, that we describe in~\cref{subsec:eval-experiment1}, aims to quantify the impact of
added decryption on NMS performance to estimate the required changes for real-world deployments.

\subsection{Implementation of Prototype}\label{subsec:eval-prototype}
We implement our approach for passive and transparent TLS decryption as a proof-of-concept prototype for the Zeek NMS~\cite{paxson1999bro}.  While some features were intentionally left out to reduce complexity, our prototype is able to successfully decrypt TLS connections and perform analysis tasks on the cleartext
application data payloads in real-world scenarios. Our prototype currently consists of
a patched Zeek version based on v3.2.3 and a simple Python daemon that runs on endpoints and
forwards key material. Our prototype only supports a single, but actually one of the most popular,
cipher suites, namely \texttt{TLS\_ECDHE\_RSA\_WITH\_AES\_256\_GCM\_SHA384}. This cipher suite was
chosen as it is commonly used in TLS 1.2 and exhibits several features of a modern cipher suite
like perfect forward secrecy. It uses the ECDHE key exchange with RSA for authentication and AES in
the Galois/Counter Mode (GCM) for symmetric encryption of payload data. GCM is favorably to our
prototype as it was designed with parallel processing in mind and allows each block to be decrypted
separately. Thus, our prototype is in worst-case also able to start the decryption in the middle of
the connection and after some initial packets could not be decrypted due to missing key material.

We \emph{obtain key material} via the SSL keylog interface as described
in~\cref{subsec:system-obtain}. We instrument the Firefox browser to write TLS pre-master secrets
and client randoms to a file monitored by our Python daemon which are then forwarded to Zeek.
Additionally, we configure Firefox to offer \texttt{TLS\_ECDHE\_RSA\_WITH\_AES\_256\_GCM\_SHA384}
as the only supported ciper suite in the TLS handshakes to ensure that our Zeek version can decrypt
the established connections.
%
The Python daemon maintains a connection to the Zeek instance via \emph{broker}, Zeek's default
communication library, and \emph{transfers new secrets to Zeek} once they appear in the file.
The daemon currently does not perform any filtering and thus does not support \emph{selective
decryption}. Instead all client randoms and pre-master secrets are forwarded to Zeek independent of
the domains involved. A complete implementation would need to match connections identifiers and
secrets to domains as this information is not directly present in the SSL keylog file alone. This
could be implemented in a browser extension or via the modification of the cryptographic library.
In the case of Firefox this is NSS.
%
Our patched Zeek version is able to receive both pre-master secrets or derived TLS session keys. In
the first case, the TLS key derivation is performed once the first packet of the matching
connection is encountered and the resulting session keys are cached for later use. Once session
keys are available, Zeek \emph{decrypts the TLS application data} and forwards the cleartext to its
internal protocol detection engine and subsequent analyzers.

\begin{figure*}
    \centering
    \includegraphics[width=.9\linewidth]{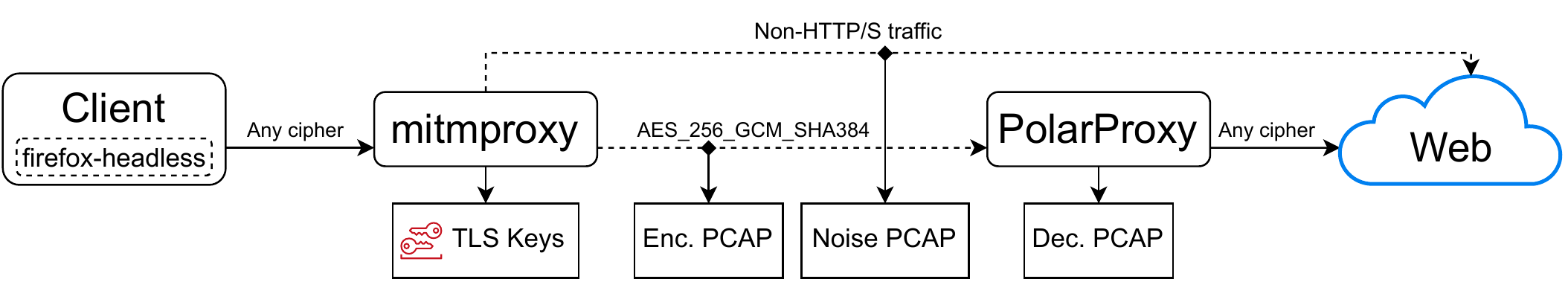}
    \caption{Setup to capture the data set to evaluate decryption overhead}\label{fig:experiment1-setup}
\end{figure*}

If our prototype encounters \emph{missing key material} for an encrypted packet, it prints a debug
message and skips decryption for this packet only. We do not perform any packet buffering directly
in Zeek which results in a partial decryption of the overall connection is the key does not arrive
in time. Usually this means that the contained HTTP/2 data cannot be fully parsed. However, packets
should typically not be buffered in the NMS itself but rather by a dedicated application in front
of it. As this is an important question for real-world deployments, we estimate the required buffer
size to avoid missing key material in our second experiment, that we describe
in~\cref{subsec:eval-experiment2}.
%
Support for TLS session resumption is currently not implemented, but can be added by linking the
respective connection identifier (Session IDs, tickets or PSK) to the received key material. The
different identifiers are encountered in varying parts of the connections and may even be
encrypted. However, the NMS retains complete visibility into all payloads (if key material is
available in time for the first packet) and can thus retrieve and store the relevant identifiers.
While it was left out in our prototype to reduce complexity, it should definitely be supported in a
full implementation, as TLS session resumption is used widely across the web especially for large
content delivery networks~(CDNs).

In summary, we build a prototype that serves as a proof-of-concept NMS and that is able to decrypt
TLS traffic in a passive manner. The largest limitation is the support of only a single ciper suite
and the missing support for TLS session resumption. Given either pre-master secrets or the
respective session keys, our patched Zeek version is able to analyze TLS-wrapped HTTP/2 traffic as
if it were captured in cleartext. The prototype is fully available on github\footnote{\patchurl}.

\subsection{Experiment I: Decryption Overhead}\label{subsec:eval-experiment1}
In this first experiment we evaluate the central performance metric for our prototype
implementation: \emph{decryption overhead},\ie{}the additional runtime added by performing TLS
decryption. This metric is important when deploying TLS decryption, as NMS machines need to be
scaled up to accommodate for the additional computation requirements. There are two interesting
scenarios to consider:
\begin{enumerate}
    \item In the first scenario, an organization is currently not deploying MitM proxies to
        terminate and decrypt TLS connections. Instead, a regular NMS is used to analyze encrypted
        traffic at the network boundary. Due to the encryption, the NMS is only able to analyze
        connection meta-data resulting in a comparatively low resource utilization.
    \item In the second scenario, the organization follows industry best-practices and deploys a MitM
        proxy that forward payloads from terminated TLS connection to a NMS. In this case, the NMS
        has full cleartext access to all connections and can perform a more sophisticated analysis,
        which results in higher resource utilization.
\end{enumerate}

In both scenarios the NMS machines need to be scaled up, as more computation will be performed.
However, in the second scenario, MitM proxies are no longer required and their previously allocated
resources can be used for monitoring. To evaluate the decryption overhead for both scenarios, we
design a capture environment that simulates both deployments.

\subsubsection{Data Set}\label{subsubsec:eval-experiment1-dataset}
\Cref{fig:experiment1-setup} shows the setup we used to capture the data set for this experiment. The
client machine runs Firefox in headless mode that visits all Alexa Top 1000 websites one after the other.
The traffic is transparently routed through \texttt{mitmproxy}~\cite{mitmproxy} to extract the TLS
pre-master secrets and force the cipher suite that our prototype implements. The outgoing traffic from
this machine is then routed differently based on the destination port. Web traffic destined for the ports
80/443 is forwarded to another host running
\texttt{PolarProxy}\footnote{https://www.netresec.com/?page=PolarProxy}, while all remaining non-HTTP/S
traffic is routed directly to the Internet. We use PolarProxy as it can export decrypted HTTP/2 traffic
directly to a pcap file, a feature that \texttt{mitmproxy} does not support. This pcap represents the
commonly deployed second scenario described above in which the NMS receives decrypted traffic from a TLS
intercepting proxy. The encrypted traffic and noise traffic is also captured and stored in different pcap
files. These pcaps represent the aforementioned first scenario in which the NMS analyzes encrypted
traffic only.

The resulting data set captured from our deployment comprises the \emph{TLS pre-master secrets}
extracted from mitmproxy and pcaps for \emph{encrypted web traffic}, \emph{other noise
traffic} as well as the \emph{decrypted HTTP/2 payloads}. All pcaps and the captures TLS pre-master secrets
are available on mega.nz\footnote{\pcapurl}.

\subsubsection{Results}\label{subsubsec:eval-experiment1-results}
By using our captured data sets, we can now compare the runtime of both, the unmodified Zeek version and our prototype.
Please note that this first experiment focuses on decryption performance and overhead alone. As such, we configure our Zeek prototype to load all TLS pre-master secrets from disk at the start of the experiment and let Zeek analyze the respective scenario pcaps. This results in all keys being instantly available, which is not comparable to real-world deployments. To address this, we analyze the impact of real-time transmission of TLS keys and their delayed arrival at the NMS in the second experiment (c.f.~\cref{subsec:eval-experiment2}).

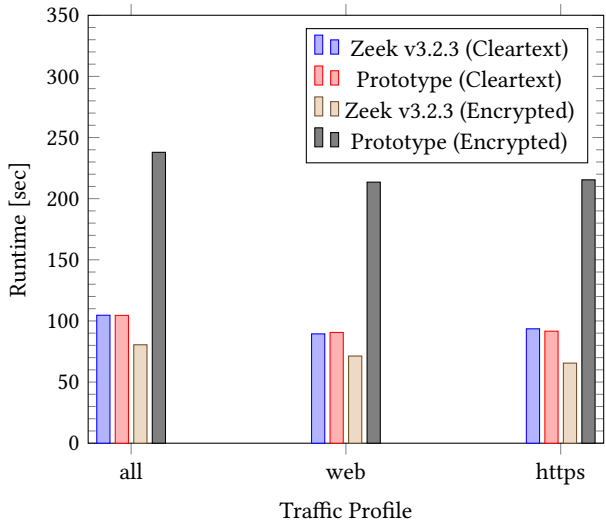
\begin{figure}
    \centering
    \pgfplotsset{compat=1.3}
    \begin{tikzpicture}
        \begin{axis}[
            bar width=5pt,
            xlabel={Traffic Profile},
            xtick=data,
            ylabel={Runtime [sec]},
            ybar,
            ymin=0,
            ymax=350,
            ytick distance=50,
            minor y tick num=4,
            symbolic x coords={all,web,https},
            legend pos=north east,
            no marks,
        ]

        \addplot plot coordinates {%
                (all, \ohalldecnormal) (web, \ohwebdecnormal) (https, \ohhttpsdecnormal)};
        \addplot plot coordinates {%
                (all, \ohalldecmod) (web, \ohwebdecmod) (https, \ohhttpsdecmod)};
        \addplot plot coordinates {%
                (all, \ohallencnormal) (web, \ohwebencnormal) (https, \ohhttpsencnormal)};
        \addplot plot coordinates {%
                (all, \ohallencmod) (web, \ohwebencmod) (https, \ohhttpsencmod)};

        \legend{%
            Zeek v3.2.3 (Cleartext),
            Prototype (Cleartext),
            Zeek v3.2.3 (Encrypted),
            Prototype (Encrypted)
        },
        \end{axis}
    \end{tikzpicture}
    \caption{Decryption overhead}\label{fig:experiment1-results}
\end{figure}

\cref{fig:experiment1-results} shows the decryption overhead by comparing the runtime of the
unmodified Zeek version and our prototype for three different traffic profiles. Each traffic
profile is defined in both \emph{decrypted} and \emph{encrypted} fashion as follows:
\first \textbf{https} contains only traffic that was originally sent via HTTPS. This includes most
websites from Alexa top 1000 minus a few that did not support TLS at all. The decrypted variant
is built from the decrypted pcap written by PolarProxy while the encrypted variants derives from
the pcap captures between mitmproxy and PolarProxy. For HTTPS both pcaps are filtered by
destination port 443. \second \textbf{web} consists of the same pcaps but removes the port filter
and consists of all HTTP and HTTPS traffic. \third \textbf{all} extends the previous traffic
profile by adding non-HTTP/S noise traffic. The decrypted variant simply merges the decrypted pcap
by PolarProxy with the noise pcap and the encrypted variant merges the pcap captured between the
proxy machines with the noise pcap.

The first two bars per traffic type show the runtime for the cleartext traffic for both the unmodified Zeek and our prototype. For all three traffic types, the runtime is nearly identical. This is expected as no decryption has to be performed and both version perform identical analysis tasks on cleartext data. These measurements thus resemble the NMS runtime in a typical MitM proxy deployment without including the runtime of the proxy itself.
The third bar shows the runtime of the unmodified Zeek version analyzing encrypted traffic. The runtime is smaller than for the cleartext traffic, as this Zeek version cannot decrypt the payloads and thus falls back to metadata only analysis. This difference is most visible form the HTTPS only data set as no noise or HTTP data is present, which would be in cleartext and thus potentially triggering complex analysis scripts. This measurement also establishes a baseline for the first scenario described in the introduction of this section: a NMS processes encrypted network traffic and no MitM proxy is deployed.
The fourth bar shows the runtime our prototype that decrypts the TLS connections and analyzes the contained payloads. The added decryption results in a comparatively large runtime overhead as exptected. Compared to the MitM baseline (cleartext traffic), we see an increase by a factor of 2.5. An administrator should scale the resources of NMS machines by this factor when switching from an MitM proxy deployment to our approach. As MitM proxies are no longer needed in this scenario their resources could be repurposed.
Compared to the first scenario (encrypted traffic), the overhead grows to about 3 times. However, this is reasonable, as no decryption was performed previously (neither by a MitM proxy nor the NMS). The previous comparison and our discussion in~\cref{subsec:eval-computation} show that deploying a MitM proxy would likely result in higher resource usage.

\subsection{Experiment II: Impact of Delay for Retrieving Keys}\label{subsec:eval-experiment2}
\begin{figure}
    \centering
    \includegraphics[width=\linewidth]{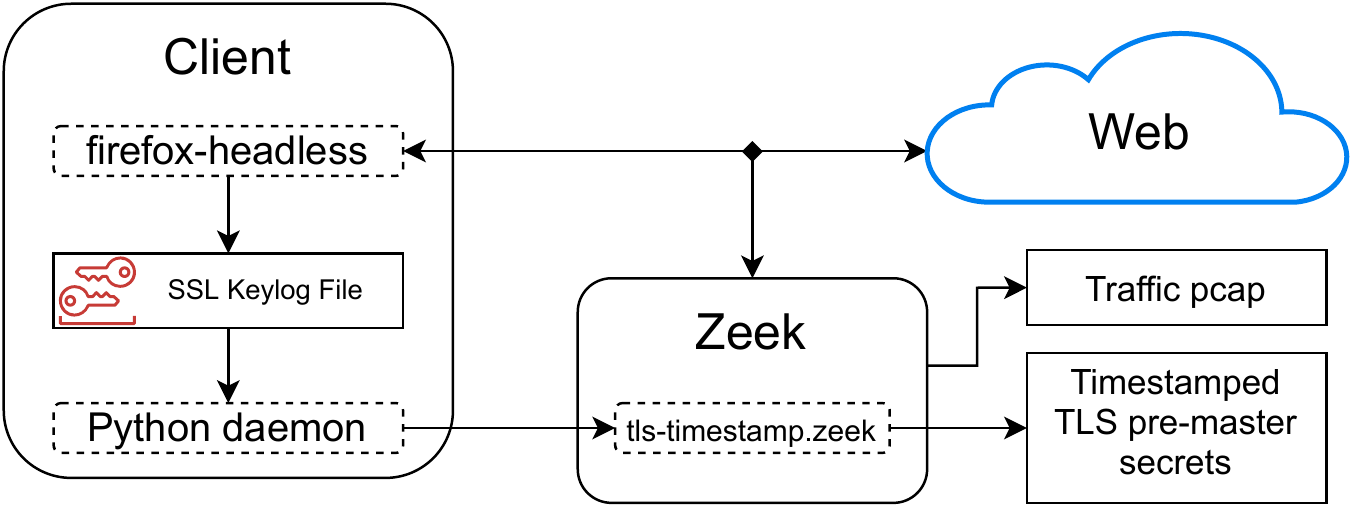}
    \caption{Setup to capture the data set to evaluate impact of key transfer latency}\label{fig:experiment2-setup}
\end{figure}

In the second experiment, we evaluate the real-world viability of our approach by examining the delay between the key material being made available on the client machine and the arrival at the NMS after transfer through the network. This is especially relevant, when considering cipher suites that require the cleartext of the previous packet when decrypting a newly arriving packet. For these ciphers
either the key material needs to be available once the first encrypted packet arrives, the NMS needs to buffer packets, or the network traffic needs to be delayed before being passed to the NMS.

\subsubsection{Data Set}\label{subsubsec:eval-experiment2-dataset}
We design a deployment to capture timestamped TLS pre-master secrets for this experiment that we show in~\cref{fig:experiment2-setup}. The client machine again runs Firefox in headless mode and is configured to only offer the prototype's supported cipher suite. We then browse the Alexa top 1,000 websites and write the SSL keylog file as usual. Our python daemon also runs on the client machine and watches the keylog file for changes via \texttt{inotify}. New pre-master secrets are immediately forwarded to Zeek via its Broker communication library. On arrival, Zeek timestamps the key material to document the arrival time. Additionally, the machine running Zeek is used as the gateway for the client host which allows Zeek to capture the traffic as a pcap.
The resulting data set comprises the \emph{traffic pcap} as well as the \emph{timestamped TLS pre-master secrets}.
It is available on mega.nz\footnote{\pcapurl}.

\subsubsection{Results}\label{subsubsec:eval-experiment2-results}
For our second experiment, the timestamped TLS pre-master secrets are again pre-loaded in our prototype that then analyzes the captured pcap file.
However, the timestamp now dictates when Zeek starts to actually use the key. For the default case, it only starts decryption once the network timestamp in the pcap is larger than the timestamp of the corresponding pre-master secret. This resembles a real-world deployment without any traffic buffering.
For simulating this traffic buffering, we have two options: \first we can alter the timestamps in the pcap file by the respective time interval while leaving the pre-master secret timestamps untouched. This closely resembles delayed traffic. \second we can use the pcap as is and simply add the desired time interval to the current network time in the comparison operation in Zeek. For simplicity we chose the second option as the delay can easily be configured in a script.
In the experiment runs, we count the number of TLS payload bytes that could be decrypted. As our  implemented cipher suite can start to decrypt from any packet in the connection, this can be anything between 0 (if the key arrives after the connection is already finished) or the full amount of payload
bytes in the connection. Additionally, we consider a connection completely decryptable if the key is available before the first encrypted payload packet is encountered. This is important for two reasons:
\first some cipher suites can only decrypt all or no packets in a connection as packet payload is used as IV in decryption of the next packet.
\second especially for HTTP/2 the first bytes of the connection are important for the NMS as they contain the client request. If they are missed, the NMS cannot analyze the protocol level.
The \emph{decryption performance} is then calculated as the ratio of decrypted bytes/connections and total bytes/connections respectively.

\begin{figure}
    \begin{tikzpicture}
        \begin{axis}[
            xlabel={Traffic Delay [ms]},
            xmin=0,
            xmax=100,
            minor x tick num=1,
            xmajorgrids,
            ylabel={Decryption Success Rate},
            ymin=0,
            ymax=1,
            mark size={3pt},
            grid style=dashed,
            enlargelimits=0.02,
            legend pos=south east,
            cycle list name=myline
        ]

        \addplot table[x=offset,y=conns,col sep=tab] {delay.tsv};
        \addplot table[x=offset,y=tls_bytes,col sep=tab] {delay.tsv};

        \legend{
            Connections,
            TLS Bytes,
        }
        \end{axis}
    \end{tikzpicture}
    \caption{Decryption success rate depending on traffic delay\label{fig:experiment2-results}}
\end{figure}
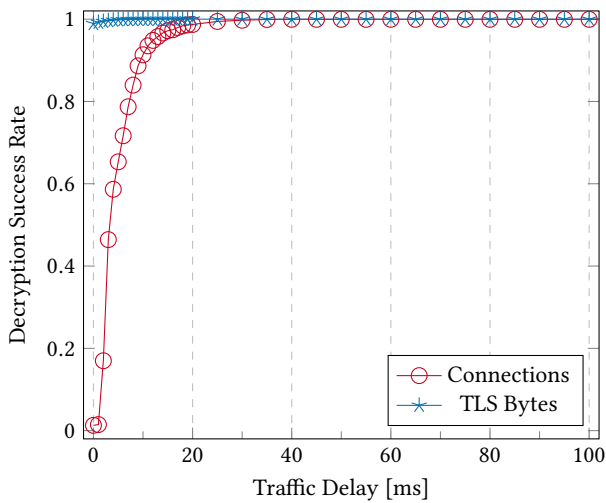

Our results are shown in~\cref{fig:experiment2-results}. The x-axis shows the simulated traffic delay in milliseconds and the y-axis the decryption success rate,\ie{}the respective percentage of bytes and connections that could be correctly decrypted. In a real-time deployment and without any delay only 1.3\% of all connections are completely decryptable. This is expected as the key transfer latency prevents the first packets from being decrypted. However, 98.9\% of all TLS payload bytes can already be decrypted. This indicates that already only few of the first bytes are missed. This is confirmed in the first few 1ms steps as the success rate for connections increases  sharply. At 5ms traffic delay already 65.3\% of all connections can be decrypted while 10ms delay results
in a success rate of 91.3\% for connections and 99.96\% of TLS bytes respectively. The success rates
steadily increase until at 40ms effectively 100\% of both bytes and connections can become decrypted.

In summary, our results from the second experiment indicate, that only a small delay (potentially as small as 40ms) is required to achieve nearly complete coverage for decryption. For real-world deployments, traffic can be sufficiently buffered before passing it to the NMS. Depending on the utilization of the tapped network link, this results in a buffer size of up to 400M for a typical 10G uplink in a medium sized network. It is important to note that this small delay is only required before passing traffic to the NMS and thus does not impact other hosts in the network.
Problems may arise if the NMS is also used as an intrusion prevention system~(IPS) and thus controls a firewall. In these scenarios, the delay may result in the IPS reacting too slow to effectively block intrusion attempts. Especially in cases like these, it might be advisable to implement the buffering as part of the TLS parsing layer of the NMS itself---which is more complex, but in turn does not increase reaction latency.

\section{Conclusion}\label{sec:conclusion}
In this paper, we present an approach to passively decrypt TLS traffic on a trusted network monitoring system~(NMS).
Our approach is based on client machines selectively forwarding TLS key material to a trusted NMS to enable
decryption without prematurely terminating the TLS connection. This provides several advantages
compared to man-in-the-middle~(MitM) proxies which are commonly deployed in enterprise networks. No additional latency is added, as no additional communication partner is involved and user retain some control about which connections can be decrypted by the NMS. Our approach can also preserve end-to-end data integrity if non-AEAD ciphers are used. In this case the NMS only receives TLS encryption keys and is thus not able to forge packets. Thus, this makes a strong case for introducing non-AEAD cipher suites to TLS 1.3 again.

We implemented our approach as a prototype for the Zeek NMS~\cite{paxson1999bro} and a python daemon that monitors extracts and forwards the keys to Zeek. Both TLS key derivation and the actual decryption is performed by openSSL, as Zeek already links to this library. While our implementation has some limitations, most notably support for only a single, widely used cipher suite, it shows promising results in both decryption overhead and real-world applicability.
Our results indicate, that TLS decryption increases the runtime by a factor of 2.5 when compared to analyzing cleartexts only. This number sounds large initially, however it is important to consider that our approach does not require a man-in-the-middle~(MitM) proxy to extract cleartext traffic. Our complexity analysis reveals that our approach requires significantly less computational resources than MitM proxies. Furthermore, we evaluated how the transfer of TLS key material impacts the decryption success rate. This is important as the NMS needs to receive the key material before the first encrypted packet is encountered to ensure full decryption of all application data packets. Our results indicate, that in our simple setup keys arrive slightly to late if traffic is not delayed before
being passed to the NMS. However, a small delay such as 40ms is enough to fully decrypt 99.99\% of all observed TLS connections.
Overall, we believe that our approach and our proof-of-concept prototype is usable for real-world analysis tasks.

While our results are promising and allow to move TLS decryption into the NMS, there are several
improvements possible as future work:
\first we plan to extend and upstream our prototype to the Zeek NMS to foster adoption in the community.
While additional cipher support is of high-priority, we also plan to integrate the client part of our
prototype with the existing Zeek Agent project~\cite{zeekagent}.
\second the current prototype only supports TLS over TCP and thus is unable to handle QUIC-based TLS that
is used by the upcoming HTTP/3 standard. However, the extension should be simple as the overall protocol
including the handshake is nearly identical.
\third an interesting variation of our concept could be clients sending the TLS key material to a trusted
third party~(TTP) as key escrow instead of the NMS while the encrypted network traffic is captured
independently. For normal operations no connections are decrypted and thus user privacy retained. In case
of an intrusion or another anomaly, the key escrow,\eg{} a worker union, could provide selected key
material to threat hunters to enable selective decryption for forensic analysis while also protecting the
privacy of uninvolved users.

\balance%
\bibliographystyle{ACM-Reference-Format}
\bibliography{wilkens21tls}


\end{document}